\documentclass[useAMS]{mn2e}
\usepackage{mnras_cite}
\usepackage{epsfig}

\def\H2{{\rm H}$_2$}

\begin{document}
\title[IRAC Photometric analysis and the Mid-IR Photometric Properties of LBGs]{IRAC Photometric Analysis and the Mid-IR Photometric Properties of Lyman Break Galaxies}\author[Magdis et. al]{G.E. Magdis$^{1}$, D. Rigopoulou$^{1}$, J.-S. Huang$^{2}$, G.G. Fazio$^{2}$, S.P. Willner$^{2}$, \newauthor M.L.N. Ashby$^{2}$\\
\vspace{0.5cm}\\
$^{1}$Department of Physics, University of Oxford, Keble Road, Oxford OX1 3RH, United Kingdom\\
$^{2}$Harvard-Smithsonian Center for Astrophysics, 60 Garden Street, Cambridge, MA 02138 
}

\maketitle

\begin{abstract}
We present photometric analysis of deep mid-infrared observations obtained by Spitzer/IRAC covering the fields Q1422+2309,
Q2233+1341, DSF2237a,b, HDFN, SSA22a,b and B20902+34, giving the number counts and 
the depths for each field. In a sample of 751 LBGs lying in those 
fields, 443, 448, 137 and 152  are identified at 3.6$\mu$m, 
4.5$\mu$m, 5.8$\mu$m, 8.0$\mu$m IRAC bands respectively, expanding their spectral energy distribution to rest-near-infrared 
and revealing that LBGs display a variety of colours. Their rest-near-infrared properties are rather inhomogeneous, ranging from
those that are bright in IRAC bands and exhibit $[R]-[3.6]>1.5$ colours to those that are faint or not detected at all in 
IRAC bands with $[R]-[3.6]<1.5$ colours and these two groups of LBGs are investigated. We compare the mid-IR colours of the LBGs 
with the colours of star-forming galaxies and we find that LBGs have colours consistent with star--foming galaxies at z$\sim$3. 
The properties of the LBGs detected in the
 8$\mu$m IRAC band (rest frame K-band) are examined separately, showing that they exhibit redder$[R]-[3.6]$ colours 
than the rest of the population 
and that although in general, a multi--wavelength study is needed to reach more secure results, IRAC 8$\mu$m band can be used as a
diagnostic tool, to separate high z, luminous AGN dominated objects from normal star-forming galaxies at z$\sim$3. 
\end{abstract}

\section{Introduction}
Observation and study of high-redshift galaxies is
essential to constrain the history of galaxy evolution and give us a systematic 
and quantitative
picture of galaxies in the early universe, an epoch of rigorous star and galaxy formation.
Large samples of high-z galaxies that have recently become available, play a key role to 
that direction and have revealed a zoo of different galaxy populations at z$\sim$3.
There are various techniques for detecting high-z galaxies involving observations in 
wavelengths that span from optical to far-IR. Among these, the three most efficient are
1)sub-mm blank field observations, 
using the sub-mm Common User Bolometer Array (SCUBA) on the James Clerk Maxwell Telescope 
(JCMT) (e.g., Hughes et al. 1998) or the Max Plank Millimeter Bolometer array (MAMBO, e.g., 
Bertoldi et al. 2000), taking advantage of the strong negative K-correction effect and revealing 
the population of the sub-mm galaxies at z$>$2 (Chapman et al. 2000, Ivison et al. 2002, 
Smail et al. 2002) and 2)the U-band-dropout technique (Steidel $\&$ Hamilton 1993, Steidel et al. 1999, 2003 
Franx et al. 2003, Daddi et al. 2004), sensitive to the presence of the 
921{\AA} break, designed to select z${\approx}$3 galaxies and revealing the population of 
the Lyman Break Galaxies (LBGs), 3) the Near-IR colour  (J$-$K $>$ 2.3 (Franx et al. 2003) and BzK ((z$-$K)$_{\rm AB}$ $-$ (B$-$z)$_{\rm AB}$$ >-$0.2) (Daddi et al. 2004)
selection for old and dusty galaxies at 2$<$z$<$3 (Distant Red Galaxies (DRGs) and BzK galaxies).

With recent deep optical and IR observations of these 3 types of galaxies, there has been considerable progress 
in understanding the relation between LBGs, SMGs, DRGs and BzK galaxies.
Chapman et al. 2005 confirmed that some of the SCUBA galaxies have rest-frame-UV colours typical of the LBGs. 
Using IRAC and MIPS observations, Huang et al. 2006
has shown that LBGs detected in MIPS 24$\mu$m band, (Infrared Luminous LBGs) 
and cold SCUBA sources share similar [8.0]$-$[24] colours, while Rigopoulou et al. 2006 suggest
 that ILLBGs and SCUBA galaxies tend to have similar stellar masses and dust amount. A possible
 scenario is one in which sub-mm galaxies and LBGs form a continuum of objects with SMGs 
representing the reddest dustier and more intensively star-forming LBGs, but further 
investigation is required to establish a more secure link between these two populations.  

LBGs constitute at the moment the largest and most well studied galaxy population at  z$\sim$3 (Steidel et al. 2003).
 Based on observations of the 
UV continuum emission, the predicted mean star formation rate of the LBGs is 20--50$M_{\odot}$/yr 
(assuming $H_{o}$$=70 Kms^{-1}Mpc^{-1}$ and $q_{o}$=0.5, Pettini et al. 2001).
 This star formation rate increases significantly if corrected for dust attenuation, 
to a mean value of $\sim$100$M_{\odot}$/yr. Correction for dust attenuation must be
 taken into account as there is clear evidence for the presence of significant amounts of 
dust in the galactic medium of LBGs (e.g., Sawicki $\&$ Yee 1998, Vijh 2003).

To investigate the amounts of dust in the LBGs, various techniques and observations spanning 
from optical to X-rays have been used. The techniques range from studies of optical
line ratios (Pettini et al. 2001) to formal fits of the overall SED of LBGs based on various 
 star formation history scenarios (e.g., Shapley et al. 2001, 2003, Papovich et al. 2001) 
and X-ray stacking studies (e.g., Nandra et al. 2002, Reddy$\&$Steidel 2004). All approaches agree
 that LBGs with higher star formation rates contain more dust (e.g., Adelberger 
$\&$ Steidel et al. 2000, Reddy et al. 2006). 

One of the most important properties of the population is the stellar mass of the galaxies. 
Stellar masses play a central role to the understanding of the evolution of the LBGs, as the co-moving 
stellar mass density at any redshift is the integral 
of the past star-forming activity and suffers fewer uncertainties than the star formation rate.
Most of the attempts to estimate the stellar content of the LBGs were based on 
ground based observations. As the observed optical band 
corresponds to rest-UV for galaxies at z$\sim$3,
 most of the light emitted from the galaxy at these wavelengths is radiated from 
young and massive OB stars. Therefore, optical observations cannot be regarded as a 
robust tool to probe the stellar mass of the galaxy as they are sensitive only to 
recent star formation episodes rather than the stellar population that has accumulated 
over the 
galaxy's life-time. Observations at longer wavelengths, i.e., rest-near-infrared (e.g. Bell \& deJong 2001, where the 
bulk of the stellar population radiates are essential to constrain the stellar content 
of the z$\sim$3 galaxy population.

With the advent of the Spitzer Space Telescope (Werner et al. 2004) we now have access to 
longer wavelengths.
The four IRAC bands on board \textsl{Spitzer} (Fazio et. al 2004) and more particularly 
the fourth band at 8.0$\mu$m allows us to observe the rest frame K-band where 
the bulk of the the stellar population of a galaxy radiates and expand the SED of the 
galaxies from rest-UV to NIR. Thus, with IRAC data in hand we have a powerful tool 
to probe in a more secure way the stellar content of the LBGs. Recent results of adding IRAC 
photometry to stellar mass estimates have been presented in e.g., Barmby et al. 2004  for z$\sim$3, 
Shapley et al. 2005 for BX/BM objects and Reddy et al. 2006  for z$\sim$2--3. 
Also, IRAC colours combined with ground based data, can enlighten the
diversity of the population revealing the existence of different sub-classes of LBGs, and provide 
an insight to their physical properties such as the energy source that powers the LBGs. 

This paper is organised as follows: Section 2 reviews the data of this study and presents the source extraction and 
the photometric analysis, as well as, the differential number counts for each field. Section 3 focus on the mid-infrared 
identifications of LBGs, while in Section 4 the rest frame near-infrared photometric properties of the LBGs are discussed. 
Finally, Section 5 summarises the main results of this study. All magnitudes appearing in this study
 are in AB magnitude system.   

\section{\textsl{Spitzer} Observations}
The data for this study have been obtained with the Infrared Array Camera (IRAC) (Fazio et al. 2004) on the 
board Spitzer Space Telescope.
 The majority of our
data are 
part of the IRAC Guaranteed Time Observation program 
(GTO, PI G. Fazio) and include the fields: Q1422+2309 (Q1422), DSF2237a,b (DSF), Q2233+1341 (Q2233), SSA22a,b (SSA22) and B20902+34 
(B0902) while data for the HDFN come from the Great Observatories Origin Deep Survey program 
(GOODS, PI M. Dickinson). IRAC 
has the capability of observing simultaneously in four, 3.6, 4.5, 5.8, and 8.0$\mu$m bands, covering one 
 5$'$x5$'$ field at 3.6$\mu$m and 5.8$\mu$m and an adjacent 5$'$x5$'$ field at 4.5$\mu$m and 8.0$\mu$m. 
All fields discussed here were covered  
 by IRAC at 3.6, 4.5, 5.8, and 8.0$\mu$m. In general there is a big
overlap of the observing part of the sky between the four bands for each field.
This overlap is maximum between the band 1 and 3 (set1), and band 2 and 4 (set2). 
while it can be small between set1 and set2 for some fields, to the point of 50$\%$. 
Field positions, dates, covered 
area and exposure time for each observation are summarised in Table1, while the expected AB magnitudes at which 
the observations for those fields reach a 5$\sigma$ point-source sensitivity limit are given in Table 2.
\begin{table*}
\centering 
\begin{tabular}{ccccc} 
\hline
Fields&RA/DEC&Date&Area (deg$^{2}$)    &Exp. Time (hrs)\\
\hline
HDFN          &12h36m49.6s +62d13m27s&2004-05-16&0.096&$\sim$95.27\\ 
\hline
DSF2237a,b    &22h39m06.7 +12d00m56s &2004-07-04&0.0173&$\sim$1.5\\  
\hline
SSA22a,b      &22h17m24.1 +00d11m32s &2006-07-12&0.125&$\sim$1.5\\       
\hline
Q1422+2309    &14h24m40.7 +23d00m19s &2004-01-10&0.024&$\sim$1.0\\
\hline
Q2233+1341    &22h36m23.3s +13d59m07s&2003-12-20&0.017&$\sim$1.0\\
\hline
B20902+34     &09h05m22.5 +34d08m22s &2005-05-07&0.017&$\sim$1.0\\
\hline
\end{tabular}

\caption{Details of the fields covered by this study}

\end{table*}

The IRAC Basic Calibrated Data (BCD) delivered by the Spitzer Science Center (SSC) include flat-field corrections, 
dark subtraction, linearity correction and flux calibration. The BCD data were further processed by our team's own refinement routines. 
This additional reduction steps include distortion corrections, pointing refinement, mosaicking and cosmic ray removal 
by sigma clipping. Finally, the fields that were observed more than one time and at different position angles, removal of 
instrumental artifacts during mosaicking was significantly facilitated.

\begin{table*}
\begin{center}
\begin{tabular}{c c c c c c c}
\hline
&\multicolumn{4}{c}{$3.6{\mu}m$}&&\\
\hline
Properties/Field & HDFN & Q1422 & Q2233 & SSA22 & DSF2237 & B0902 \\
\hline
${5\sigma}$ Limit$^{1}$&25.5&22.9&22.9&23.2&23.2&22.9\\
Mag. Limit$^{2}$&27.8&24.8&24.7&25.1&25.2&24.7\\
Sources Extracted$^{3}$&25615&4105&2615&28219&3403&3112\\
Completeness 50$^{4}${\%} &24.9&23.1&23.1&23.7&24.0&23.4\\
 \hline
 &\multicolumn{4}{c}{$4.5{\mu}m$}&&\\
 \hline
 $5\sigma$ Limit&25.2&22.9&22.9&23.1&23.1&22.9\\
 Mag. Limits&27.5&24.5&24.5&24.9&25.1&24.5\\
 Sources Extracted&24531&3994&2808&28253&3846&3005 \\
 Completeness 50{\%} &24.9&23.1&23.1&23.7&23.7&23.4 \\
 \hline
 &\multicolumn{4}{c}{$5.8{\mu}m$}&&\\
 \hline
 $5\sigma$ Limit&23.9&21.5&21.5&22.2&22.2&21.5\\
 Mag. Limits&26.1&22.6&22.5&23.1&23.1&22.6\\ 
 Sources Extracted&15527&2063&1719&16356&1818&1219\\
 Completeness 50{\%}&23.4&21.6&21.9&22.2&22.6&21.6\\
 \hline
 &\multicolumn{4}{c}{$8.0{\mu}m$}&&\\
 \hline
 $5\sigma$ Limit&23.8&21.4&21.4&22.2&22.2&21.4\\
 Mag. Limits&26.1&22.6&22.4&23.1&23.1&22.5\\
 Sources Extracted&14315&1658&1172&14192&1392&1119\\
 Completeness 50{\%}&23.4&21.6&21.9&22.2&22.2&21.6 \\
 \hline
\end{tabular}         
\end{center}    
\caption{Details on the source extraction and photometry in each field and each IRAC band}
\footnote{1}{Expected AB magnitudes at which the observations reach a $5{\sigma}$ point-source sensitivity.}\\
\footnote{2}{Magnitude limits for source extraction.}\\
\footnote{3}{Number of extracted sources.}\\
\footnote{4}{Magnitude at which the completeness of the source extraction reaches 50$\%$.}\\
\end{table*}

\subsection{Source extraction}
Source extraction was based on fitting point spread functions (PSFs) in each field. 
Because the depth of the observations varies from field to field and the sensitivity of IRAC
 drops significantly for the 5.8 and 8.0$\mu$m bands, different PSFs were chosen for each field and each band. The fact
that those fields are very crowded combined with the small number of stars in those fields, made the selection of the PSFs 
 very challenging. The situation was more complicated for bands 3 and 4 where
 the noise increases significantly. To accurately compute the best PSF, we used as many point sources located throughout 
the IRAC images as possible. The construction of PSFs was iterative taking into account not only isolated stars, 
but also stars with faint objects close to them. To produce the optimum
 PSF we developed an IDL code that uses the stars we select as appropriate for PSFs, 
subtracts objects that are blended or near to them, and creates 
a $''$clean$''$ average PSF. 

With PSFs derived as above, we performed source extraction using STARFIND. The final PSF was fitted in the image, searching
the input images
 for local density maxima with half-widths at half-maxima (HWHM) of the PSF's hwhm, and peak amplitudes 
greater than a given threshold above the local background. With a FWHM of the PSF of 1.8$''$ to 2.2$''$, virtually all
 objects are unresolved. The detection limits varied from field to field and from band to band according 
to the depth of the observation and the sensitivity of the observed band.  The source extraction for 
each field and each IRAC band was
repeated with slightly varied parameters till the optimum
source 
extraction was achieved
(the results of each run were inspected by eye to 
check if the extraction was shallow resulting in
missing real sources, or if we had gone to deep and noise was picked 
as sources).
The threshold was highly dependent 
on the exposure time of the observations. 
For fields of equal exposure time (i.e., depth) the same detection 
thresholds were used, while for deeper observations the threshold was lower. The limiting magnitudes 
of the performed source extraction and the number 
of objects detected in each field and each IRAC band are summarised in Table 2.

\subsection{Completeness and Photometry}
The uncertainties and the completeness of the source extraction in the IRAC bands magnitudes 
were estimated from an analysis employing Monte-Carlo simulations. In average 6000 artificial sources were added to each image, and were
extracted in the
same manner as the real sources. The dispersion between input and recovered 
magnitudes and number of objects provides a secure estimation of the detection rate and the photometric error in each magnitude bin. 
The incompleteness curve for each individual field and each IRAC band are illustrated in Figure 1 while the depth at which 
the completeness of our source extracting method reaches 
50$\%$ are summarised in Table 2. The incompleteness in all four IRAC bands shows the usual rapid 
increase near the magnitude limit of the images and declines sharply near the faint 
limits. Furthermore,
 there is  significant improvement in fields with larger exposure times, with the completeness of HDFN falling to 50$\%$
 at a magnitude of $\sim$25 at 3.6$\mu$m, while for SSA22 at a magnitude of $\sim$23.9.

\begin{figure}
\centering
\includegraphics[width=10cm,height=9cm,angle=-90]{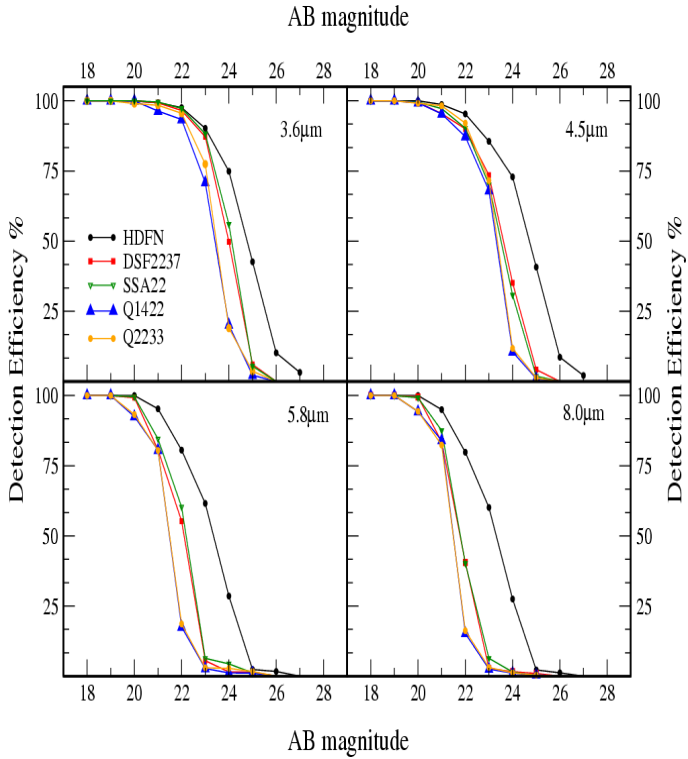}\\
\caption{\small{The incompleteness curve for all fields and IRAC bands (B0902 is excluded for presentation purposes, but it follows the
incompleteness curve of the fields with equal depth, i.e., Q1422 and Q2233). Each field in indicated by different symbol and colour. 
The incompleteness in the four IRAC bands shows the usual rapid increase near the magnitude limit of the images and 
declines sharply near the faint 
limits. Furthermore, there is  significant improvement in fields with 
larger exposure time, with the completeness of HDFN falling to 50$5$
 at a magnitude of $\sim$25 at 3.6$\mu$m,  while for SSA22 at a magnitude of $\sim$23.9}}
\label{fig:F2}
\end{figure}

For the photometric analysis, PSF fitting and aperture photometry using 
a 3$''$, 4$''$ 5$''$ and 6$''$ 
diameter 
aperture was performed. The aperture fluxes in each band were 
subsequently corrected to total fluxes using known PSF growth curves from Fazio et. al. 2004 
and Huang et al. 2005. Figure 2  shows the residual of the 4$''$ minus 
the 3$''$ diameter aperture photometry over the 3$''$ diameter aperture photometry for IRAC bands of fields HDFN and DSF.
 The strong 
concentration of the residual around zero indicates that the applied aperture 
corrections are correct and within the uncertainty of our photometry. This good agreement also holds between psf and aperture 
 photometry.The magnitude-error relation 
is shown in Figure 3 for three fields, HDFN, Q1422 and SSA22. We choose to plot these 
three fields as they cover the whole range of depths of our observations. DSF has comparable photometric error bars with SSA22,
while the magnitude-error relation for both 
Q2233 and B0902 follows that of Q1422. 
\begin{figure}
\centering
\includegraphics[width=8cm,height=9cm]{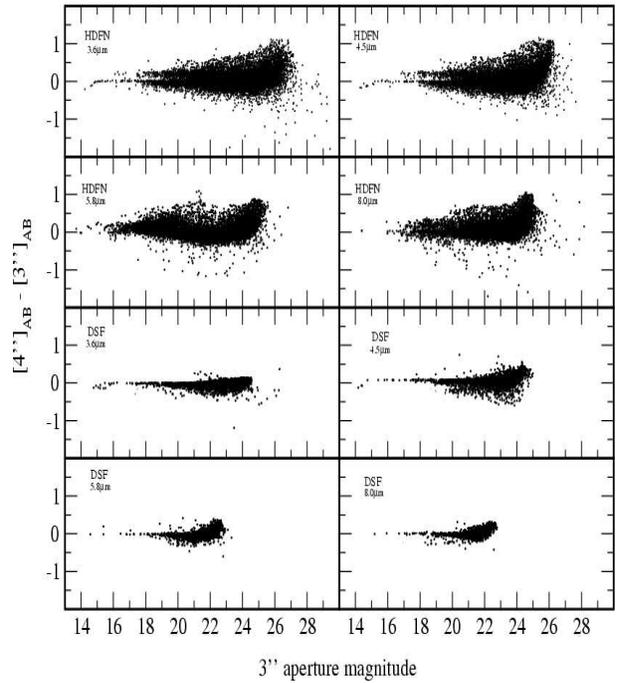}\\
\caption{\small{[4$''$]-[3$''$] versus [3$''$] magnitude for the objects of the fields HDFN and DSF. The strong 
concentration of the residual around zero, indicates that the applied aperture 
corrections are correct.}}
\label{fig:F1}
\end{figure}

\begin{figure}
\centering
\includegraphics[width=10cm,height=9cm,angle=-90]{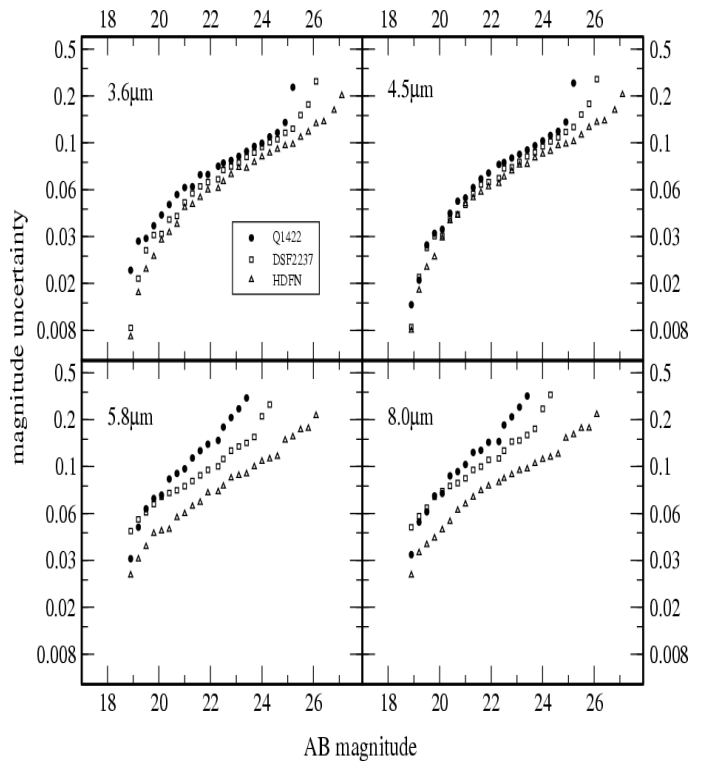}\\
\caption{\small{The magnitude-error relation for three fields, HDFN (triangles), Q1422(filled circles) and SSA22 (squares). It should be 
note that the y axis is in log scale for comparison purposes. DSF has comparable photometric error bars with SSA22,
while the magnitude-error relation for both 
Q2233 and B0902 follows that of Q1422.}}
\label{fig:F3}
\end{figure}

The four catalogues for the IRAC wavelengths were matched by position to create a single master catalogue 
of IRAC sources for each field. 
While the entire 
catalogues will be presented in a forthcoming paper anyone interested in those should 
contact the author.
 \begin{figure*}
\centering
\includegraphics[width=15cm,height=17cm,angle=-90]{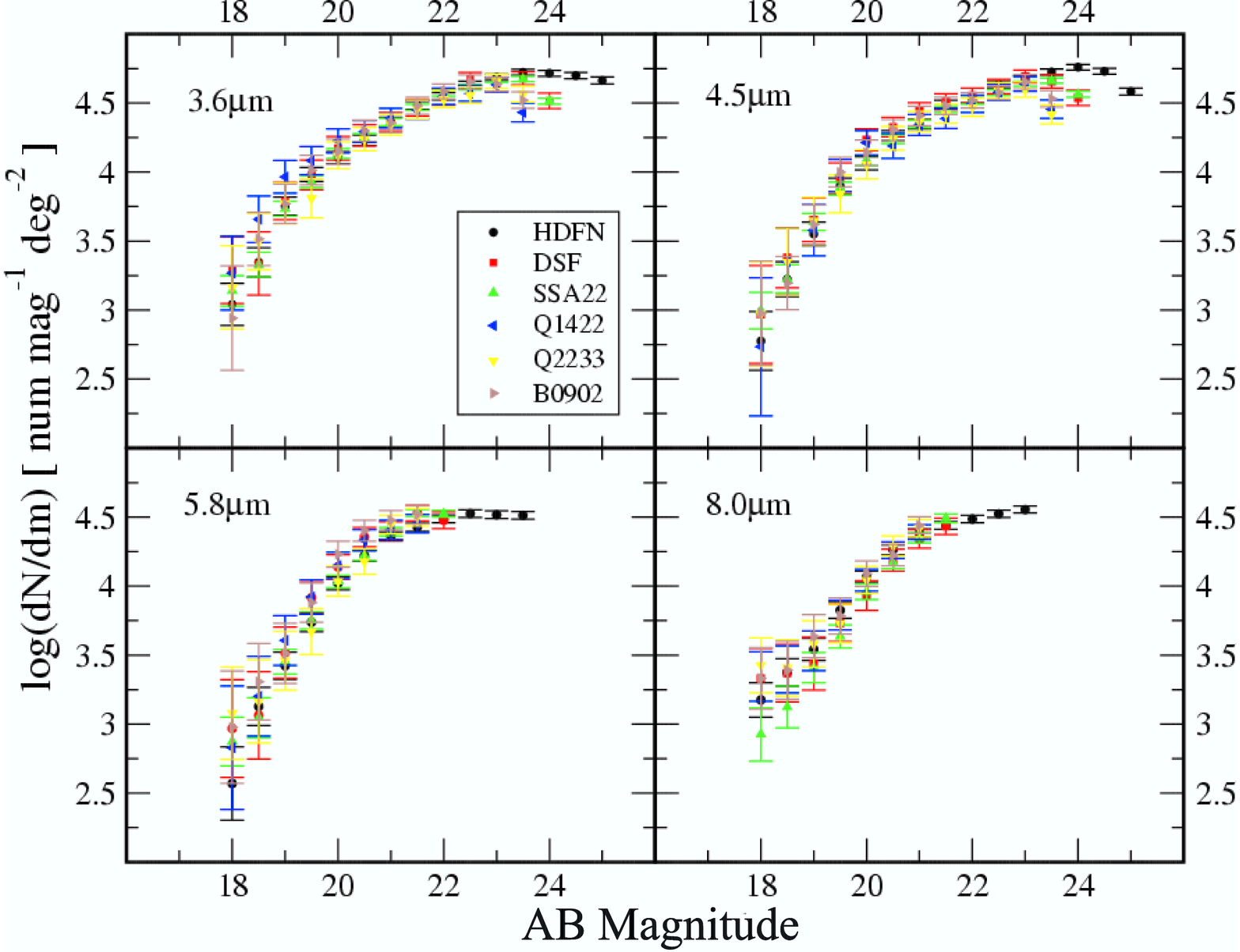}\\
\caption{\small{Differential number counts in the four IRAC bandpasses including all sources, stars and galaxies. Counts in the six
fields  are indicated by different colours and symbols. We plot the number counts only for magnitudes where the completeness is $>$ 50$\%$ There is an excellent agreement in the bright end for all the fields and in the
faint end for observations of equal exposure time (i.e., depth). The error bars represent the Poisson noise.}}
\label{fig:F4}
\end{figure*}
\subsection{Number counts}

The observations span 2.5 years and a factor of 95 in exposure time (i.e., depth).
 A reliable indicator that all fields were treated in the same manner are  the number counts. 
Figure 4  shows the differential number counts in the four IRAC bandpasses including all sources, stars, and galaxies in each field 
versus the magnitude bins. We plot the number counts for magnitudes at which the completeness is $>$ 50$\%$, 
while the uncertainty in the number counts is based on Poisson noise. There is an excellent agreement 
between the number counts of the fields
in the bright end and this agreement stands in the faint end 
between fields of equal exposure time. As expected 
the number counts peak in fainter magnitudes for
 fields with larger exposure times, while for fields of similar depth 
(Q2233, B2090, Q1422 and DSF2233, SSA22) the turning point is equal. It should be noted that the number counts of HDFN, despite the 
95 hours of integration, appear to peak (completeness $\sim$ 50$\%$) only 1.5--2 AB magnitude 
deeper than the fields with 1.5 hours of exposure time. 
On the other hand, the large difference in the exposure time between HDFN and the rest fields, becomes very significant for the 
number counts in magnitudes where the completeness is 50$\%$, as for shallower fields the number counts
 decline very steeply, compared to HDFN.

In Figure 5 we compare the differential number counts of the Extended Groth Strip (EGS) by Fazio et al. 
2004, Chandra Deep Field South (CDFS) area by Franceschini et al. 2006 (only for the 3.6$\mu$m band) and of HDFN 
derived by this study. The exposure time for the EGS and CDFS is $\sim$1.5h and $\sim$46h respectively. There is a very good agreement 
between the results with an excess in the faint end for our data, which is due
the very deep observations of HDFN.

\begin{figure*}
\centering
\includegraphics[width=17cm,height=15cm]{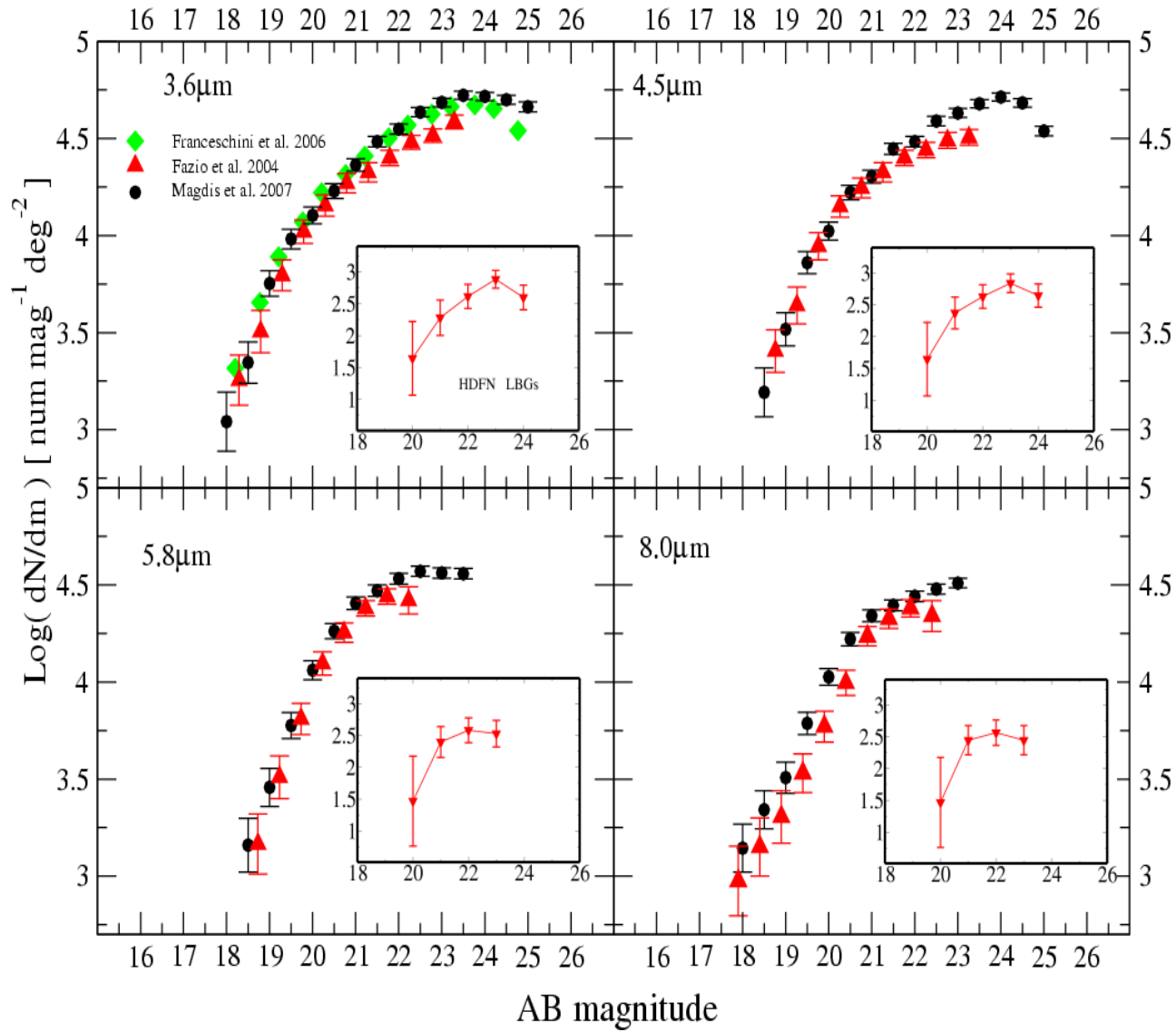}\\
\caption{\small{Comparison between differential number counts in the four IRAC bandpasses for HDFN (black circles) from this study,
 EGS (red triangles) from Fazio et al. 2004 and CDFS  (green diamonds) from Franceschini et al. 2006 (only for $3.6\mu$m). The enclosed plots show the number counts of the detected LBGs in HDFN.}}
\label{fig:F5}
\end{figure*}

\section{Mid Infrared Identification of LBGs}
Steidel et al. 2003 published a catalogue of 1261 LBGs in the six fields while our 
observations covered 751 LBGs in at least one IRAC waveband. 
The sample of 751 LBGs, consists of three categories of objects. 
Those that 
have confirmed spectroscopic 
redshift (through follow up ground based optical/near-infrared spectroscopy, Steidel et al. 2003) 
and are identified as galaxies at
z$\sim$3 (LBGs-z) or classified as AGN/QSO and, those that do not have spectroscopic redshifts.
In total, 321 LBGs-z, 12 AGN/QSO and 435 LBGs without spectroscopic redshift are covered. Out of these, 625 were covered by IRAC 
at all four wavelengths constituting our main LBG sample, while
 an additional 50 LBGs were covered at [3.6] and [5.8]$\mu$m and 93 at [4.5] and [8.0]$\mu$m. Those additional LBGs were added to 
our statistical analysis when appropriate. To identify LBGs in the IRAC images 
Steidel's catalogue was matched to 
the IRAC source lists. The Spitzer astrometry is aligned to the ESO Imaging Survey with a typical accuracy of 
0.5$''$ (Arnouts et al. 2001). 

We searched for counterparts within a 1$''$  
diameter separation centred on the optical position.
As the typical size for an LBG is 1$''$ in most cases the LBGs were 
clearly identified. Given the depth of the IRAC images, some associations with LBGs are likely to be spurious. The number of objects 
with surface density n located within a distance d from a random position on 
the sky is given by  S$={\pi}nd^{2}$ (e.g. Lilly et al. 1999). The surface density n(m) for each field at each magnitude 
is given by the differential number counts as discussed in the previous section, 
and the value d was set to 1 since for the identification of LBGs in IRAC images, a radius of 1$''$ was used for searching for
near-infrared counterpart. The number of detected LBGs in each magnitude bin was then calculated. We applied the mathematic formula
 described above and derived the expected spurious objects for each of the three categories of fields and each IRAC bandpass. 
The results are shown in Figure 7, where the magnitude distribution of the detected LBGs  
is over-plotted with that of the expected spurious objects. 
The high ratio of detected LBGs over the expected spurious objects makes the 
majority of our identifications secure. 

\begin{figure}
\centering
\includegraphics[width=9cm,height=8cm,angle=-90]{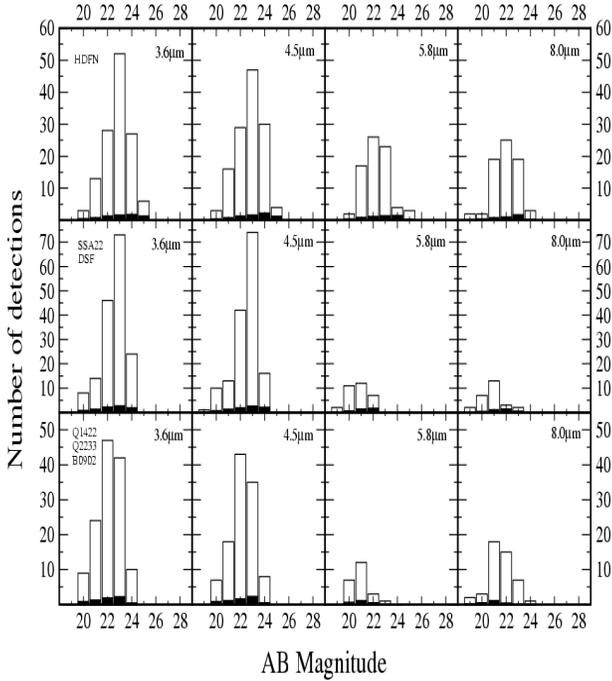}\\
\caption{\small{Magnitude distribution of detected (empty bars) and expected spurious detections for each IRAC bandpass. The fields are grouped into three sets according to the depth of their observation. The top panel shows the distribution for the HDFN, the middle for the fields SSA22 and DSF while the bottom for Q1422,Q2233 and B0902.}}
\label{fig:F6}
\end{figure} 

The detection rate of the LBGs is highly dependent on the depth of the observation, with fields of equal depth having equal 
detection rates. Therefore, the 6 fields are divided in three groups according to their depth. The first category includes 
only the HDFN as it is the field with the largest exposure time. In the second category the fields of intermediate 
depth, i.e, SSA22 and DSF are included, while the shallowest fields, 
Q1422, Q2233, and B0902 constitute the third category. Figure 6 shows the 
detection rate of LBGs at each of these categories and for all IRAC bandpasses. Apart from the large difference in the detection 
rate among several fields, we  note  the large decline in the detection rate 
at 5.8$\mu$m and 8.0$\mu$m bands compared to 3.6$\mu$m and 4.5$\mu$m for images of the same field. 
For example, in HDFN the detection rate reaches 90$\%$ for 
the first two IRAC bands while it drops to 50$\%$ for 5.8$\mu$m and 8.0$\mu$m. Table 3 summarises the LBGs covered/detected in each 
IRAC band while The total number of detections in each group of fields 
and each IRAC band are given in Table 4.

\begin{table}
\caption{LBGs detected/covered in all fields and each IRAC Band}
\begin{center}
\begin{tabular}{ccccccccc} 
\hline
Band/Type&\multicolumn{2}{c}{LBG}&\multicolumn{2}{c}{LBG-z}&\multicolumn{2}{c}{LBG-non}&\multicolumn{2}{c}{AGN}\\
\cline{2-9}
 &D&C&D&C&D&C&D&C\\
\hline

3.6$\mu$m & 443&658 & 192&263 & 241&385 & 10&10\\
4.5$\mu$m & 448&708 & 195&289 &243&409 & 10&10 \\
5.8$\mu$m & 137&658 & 54&263 & 75&385 & 8&10\\
8.0$\mu$m & 152&708 & 66&289 &77&409 & 9&10 \\

\hline
\end{tabular}
\end{center}
\label{Table.3}
\end{table}

\begin{figure}
\centering
\includegraphics[width=6cm,height=8cm,angle=-90]{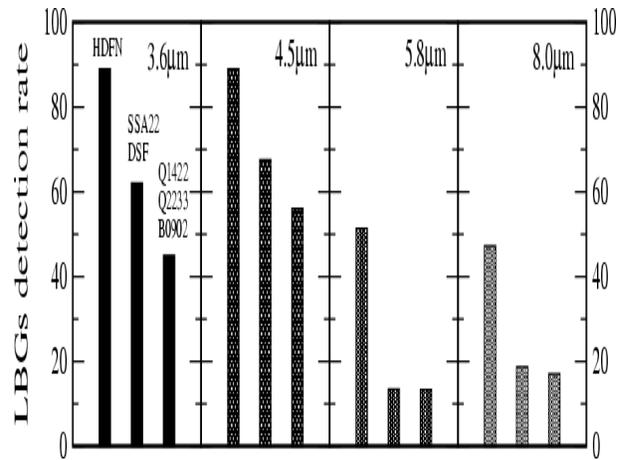}\\
\caption{\small{Detection rate of LBGs in the several fields and at each IRAC band. The fields are grouped according to the depth, 
with HDFN being the deepest field, SSA22 and DSF having intermediate depth and Q1422, 
Q2233 and B0902 being the fields with the least exposure time. The detection rate is strongly dependent on the depth of the observations.}}
\label{fig:F4}
\end{figure}

\begin{table}
\caption{Detected LBGs in fields categorised by depth, for all IRAC bandpasses.}
\begin{center}
\begin{tabular}{cccc} 
\hline
Band/Type&HDFN & SSA22 \& DSF2237 & Q1422,Q2233 \\
&&&\& B0902 \\
\hline

3.6$\mu$m & 131&170&142\\
4.5$\mu$m & 130&199&120\\
5.8$\mu$m & 75&32&30\\
8.0$\mu$m & 69&37&47\\
\hline
\end{tabular}
\end{center}
\label{Table.4}
\end{table}

    
To examine the rest-near-infrared photometric properties of the LBGs in a more complete way, 
in this paper we will focus on the 625 LBGs
that have been covered from all 4 IRAC bands.
Out of these 625 LBGs about 425 are detected 
with IRAC at 3.6$\mu$m, 401 at 4.5$\mu$m, 
136 at 5.8$\mu$m, and 149 at 8.0$\mu$m. Of these, 258 are LBGs-z, 8 
are classified as 
AGN/QSO and 359 do not have spectroscopic redshift.
In Table 5 we summarise the covered/detected LBGs that overlap between the IRAC bands.
\begin{table}
\caption{Detected/covered LBGs that are covered by all four IRAC bands}
\begin{center}
\begin{tabular}{ccccccccc}
\hline
Band/Type&\multicolumn{2}{c}{LBG}&\multicolumn{2}{c}{LBG-z}&\multicolumn{2}{c}{LBG-non}&\multicolumn{2}{c}{AGN}\\
\cline{2-9}
 &D&C&D&C&D&C&D&C\\
\hline
3.6$\mu$m & 417&615 & 183&248 & 226&359 & 8&8\\
4.5$\mu$m & 394&615 & 173&248 & 213&359 & 8&8\\
5.8$\mu$m & 129&615 & 51&248 & 71&359 & 7&8\\
8.0$\mu$m & 143&615 & 65&248 & 70&359 & 8&8\\

\hline
\end{tabular}
\end{center}
\label{Table.4}
\end{table}

\section{The photometric properties of LBGs}
\subsection{The Spectral Energy Distribution of LBGs}

With the \textsl{Spitzer} IRAC data we extend the spectral energy distribution of the LBGs to rest frame near-infrared and 
 improve dramatically our understanding of the nature of LBGs. Figure 8 shows the rest-UV/optical/near-infrared SEDs 
of all LBGs of the current sample with confirmed spectroscopic redshift, while the enclosed plot shows the SEDs of the 
LBGs-z detected in HDFN.  
UV/optical data are obtained 
from Steidel et al. 2003,
while IRAC data come from the present work. While the rest-UV/optical 
show little variation (2--3 magnitudes),
the rest frame near infrared colour spread over 6 magnitudes. The addition of IRAC bands reveals for the 
first time that LBGs display a variety of colours and their rest-near-infrared properties are rather inhomogeneous, 
ranging from : 
\begin{itemize}
\item Those that are bright in IRAC bands and exhibit $R-[3.6]>1.5$ colours. 
Their SEDs are rising steeply towards
    longer wavelengths and 
    based on their R$-$[3.6] we call them $''$\textsl{red}$''$ LBGs-z, to
\item Those that are faint or not detected at all in IRAC bands with 
    $R-[3.6]<1.5$ colour. Their SEDs are rather flat from the
far-UV to the NIR with marginal IRAC detections and as they exhibit bluer 
R$-$[3.6] colours we call them $''$\textsl{blue}$''$ LBGs-z.

\end{itemize}

\begin{figure*}
\centering
\includegraphics[width=18cm,height=16cm]{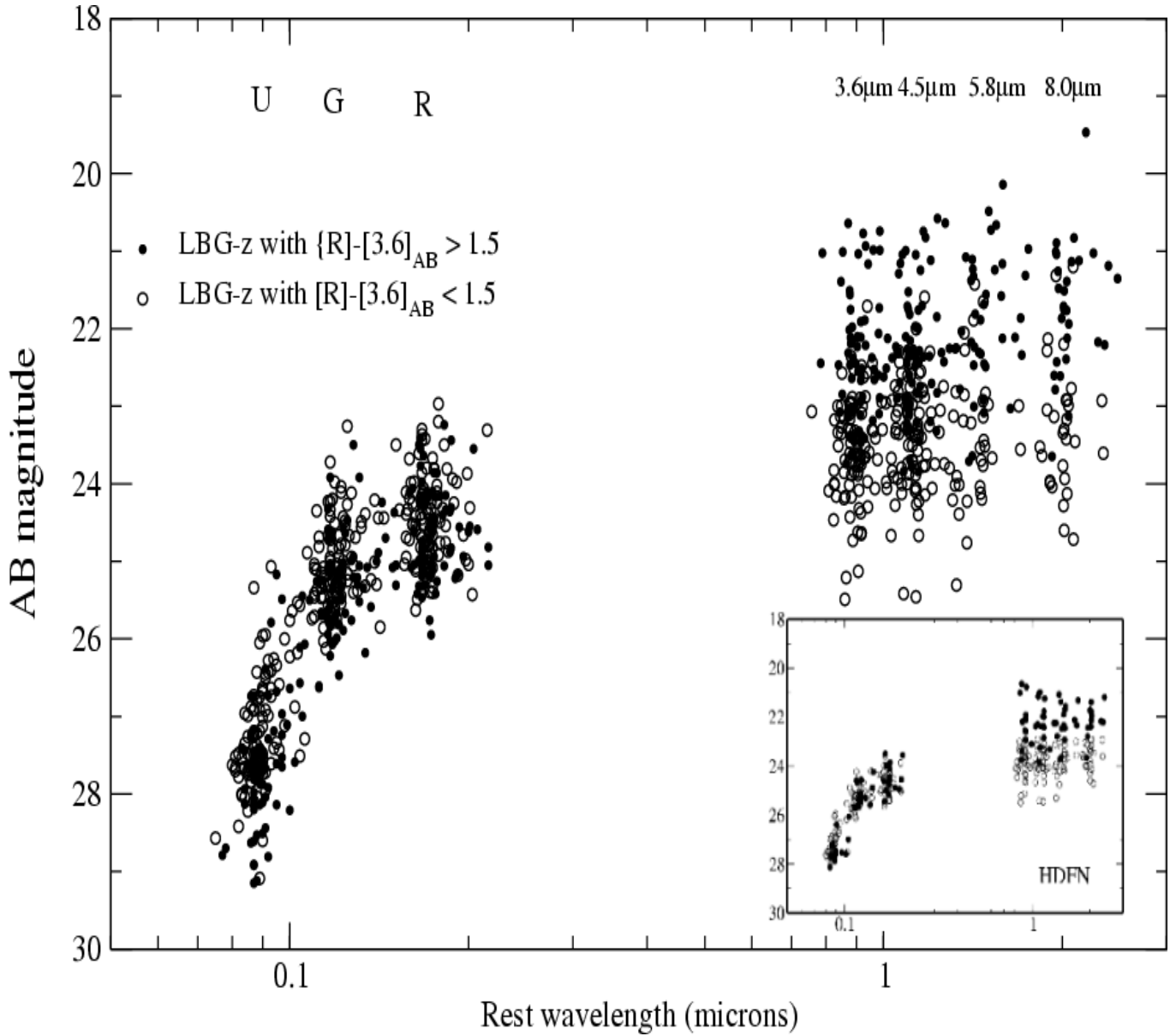}\\
\caption{\small{Spectral energy distribution for LBGs with spectroscopic redshifts and classified as galaxies. Empty circle represent LBGs with $R-[3.6]<1.5$, while filled circles represent LBGs with $R-[3.6]>1.5$, revealing that the rest-near-infrared of the population displays a wide range of colours. The enclosed plot shows the same information but only for LBGs in HDFN}}. 
\label{fig:F8}
\end{figure*}

Out of the whole sample, 3$\%$ of LBGs display R $-$ [3.6] $>$ 4, similar to the extremely red objects 
discussed by e.g. Wilson et al. 2004.

To avoid conclusions driven by selection effects and depth variance between the several fields, the sample coming from the HDFN was
 separately investigated from the other fields. The comparison of the two samples provides a simple way 
to understand the impact that the depth
 of the observation has in our sample, and therefore derive more secure global interpretations.

For LBGs with R$-$[3.6]$>$ 1.5 in HDFN, the median value (taking into account upper limits) of [3.6] and R$-$[3.6] is 22.15$\pm$0.078
 and 2.31 $\pm$0.125, respectively. For those with R$-$[3.6]$<$1.5 we get median values of 23.84$\pm$0.141 and 0.898$\pm$0.227 
while for the whole sample the derived values are 23.39$\pm$0.121 and 1.212$\pm$0.201. Kolomogorov-Smirnov test (K--S test) showed 
that the maximum difference between the cumulative distributions, D, for the [3.6] values of 
the two samples  
is 0.77 with a corresponding P-value of 0.00014, suggesting a significant difference in their IRAC 3.6$\mu$m colours with 
the $''$red$''$ LBGs being significantly brighter.

On the other hand the median value of [3.6] and R$-$[3.6] for the LBGs in the shallower fields 
is 22.51$\pm$0.095 and 1.915$\pm$0.117 for those with R$-$[3.6]$>$1.5, 23.33$\pm$0.120 and  0.975$\pm$0.214 those 
with R$-$[3.6]$<$1.5 while for the whole sample the derived median values are 23.02$\pm$0.098 and 1.372$\pm$0.195. K--S test showed again that there
is a significant difference between the [3.6] colours of 
the $''$blue$''$ and $''$red$''$ LBGs, in agreement with the results derived from
 LBGs in HDFN.  

Figure 9 shows the average
 SEDs of these two groups of LBGs-z and that of the AGNs. The selected LBGs for this plot are detected in all four IRAC bands, 
have a similar redshift (2.9$<$z$<$3.1) and are all drawn from the LBG sample of HDFN. 
The addition of the IRAC data (i.e, rest-near-infrared for our sample) reveals the difference in the SEDs of these three categories
 of objects becomes, implying the two groups of LBGs-z do not share the same properties. Further investigation of the
 physical properties of these two groups employing stellar synthesis population models, will follow in a forthcoming paper

Although the large difference in the depth of the several observations doesn't seem to affect the derived properties of the population, 
it can affect the propotion of $''$red$''$--$''$blue$''$ LBGs in each sample. 
Figure 10 shows the R$-$[3.6] colours distribution of the detected LBGs for HDFN and for the rest of the fields. 
While the distribution for LBGs with R$-$[3.6]$>$1 
 is similar for the two samples, there is an excess of LBGs with R$-$[3.6] $<$ 1 in the HDFN. The median value of [3.6] 
of the LBGs in HDFN with R$-$[3.6]$<$1 is 24.31$\pm$0.16, but as shown in Figure 6 the detection rate at the shallower 
fields is on average 
half of that in the HDFN affecting mainly the detection of faint LBGs. This can be easily understood from Figure 7. In all but HDFN 
fields, there are very few or zero detections at magnitudes fainter than 24. It is therefore expected that LBGs 
with R$-$[3.6] $<$ 1.5 are 
under-represented in the sample of the shallow fields. To quantify this under-detection of $''$blue$''$ LBGs in the shallow fields 
we  compare the fraction of population with R$-$[3.6] $<$ 1 
(where the depth of the observation becomes significant) between the two samples. 
LBGs with R$-$[3.6] $<$ 1.0 in HDFN accounts for the $\sim$31.3$\%$ of the 
total detected LBGs while in the shallower fields this fraction drops to $\sim$17.7$\%$. Therefore, a simple assumption would be that 
$''$blue$''$ LBGs in the shallower fields are under-represented by $\sim$14$\%$, that corresponds to $\sim$45 missing
 $''$blue$''$ LBGs. On the other hand, LBGs with R$-$[3.6] $>$ 1.5 have similar mean values of 
[3.6] in all fields (22.15 for HDF and 22.51 for the rest), bright enough even for the shallow fields 
to have at least comparable detection efficiency with that of HDFN.  
The above discussion is clearly demonstrated in Figure 11 where we present the [3.6] magnitude distribution 
of LBGs with R$-$[3.6] $>$ 1.5 
and R$-$[3.6] $<$ 1.5,  for the two individual samples as well as for whole sample. 
\begin{figure}
\centering
\includegraphics[width=8cm,height=7cm,angle=-90]{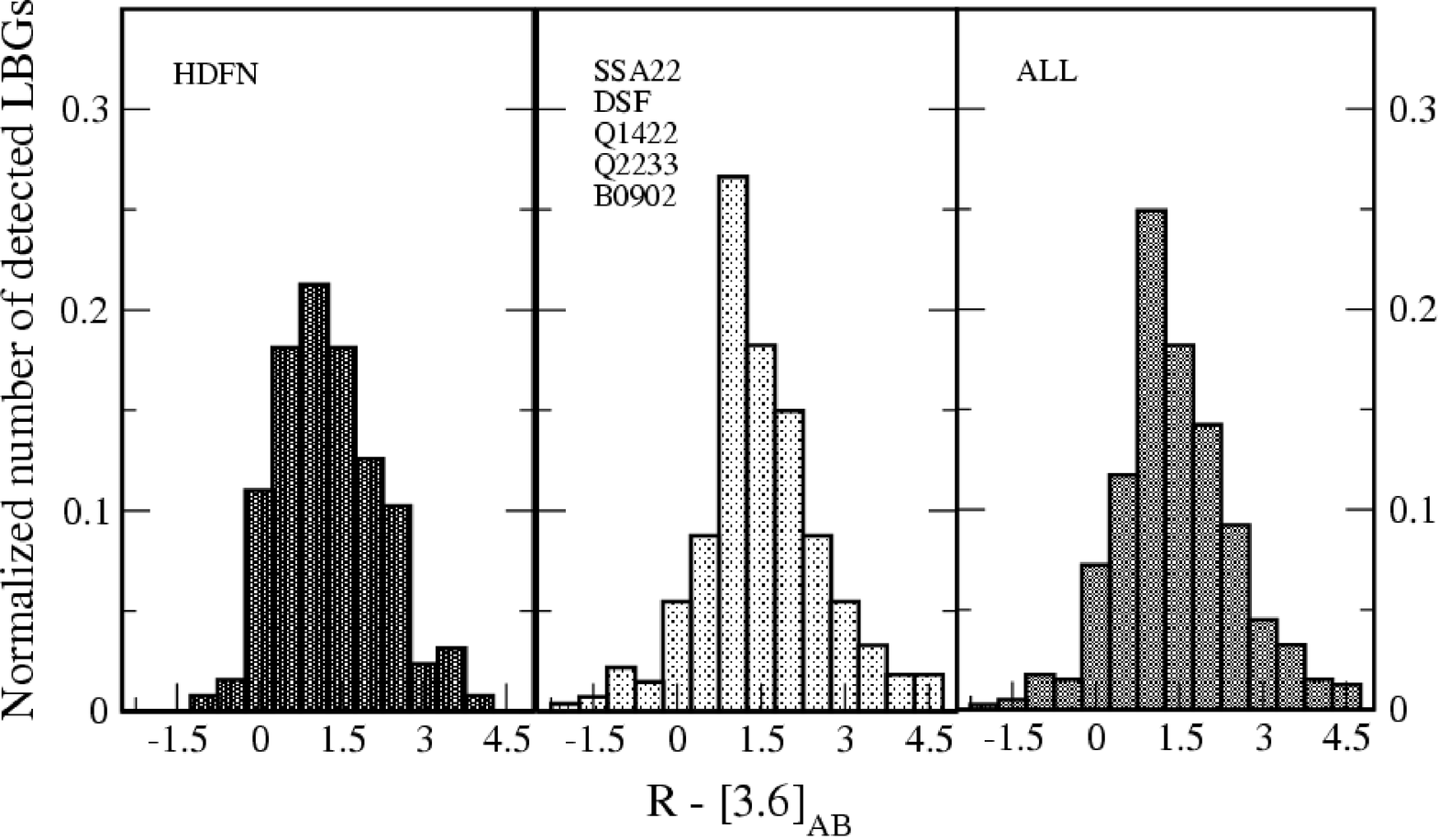}\\
\caption{\small{R$-$[3.6] colour distribution of detected LBGs, normalised to the total number of detection for each set of fields. 
The left panel describes the distribution for HDFN, the middle for the rest of the fields and the right for the whole sample of LBGs.}}
\label{fig:F9}
\end{figure}

\begin{figure}
\centering
\includegraphics[width=8cm,height=7cm,angle=-90]{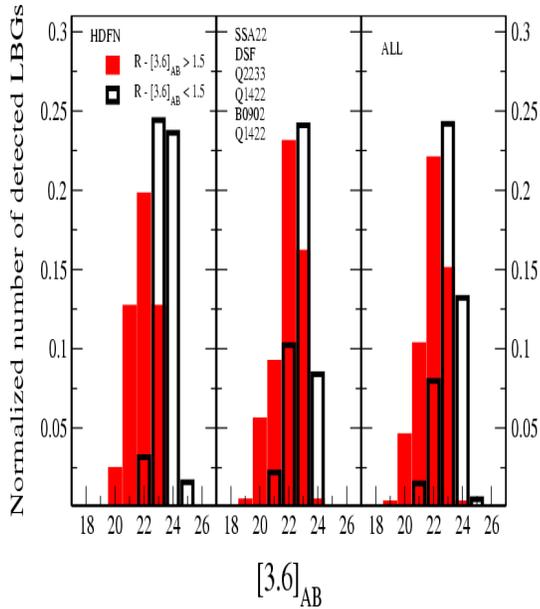}\\
\caption{\small{Normalised [3.6] magnitude  distribution of detected LBGs, for LBGs with R$-$[3.6]$>$1.5 (red bars) and R$-$[3.6]$<$1.5 (black bars). The left panel describes the distribution for HDFN, the middle for the rest of the fields and the right for the whole sample of LBGs.}}
\label{fig:F10}
\end{figure}

\begin{figure}
\centering
\includegraphics[width=10cm,height=8cm,angle=-90]{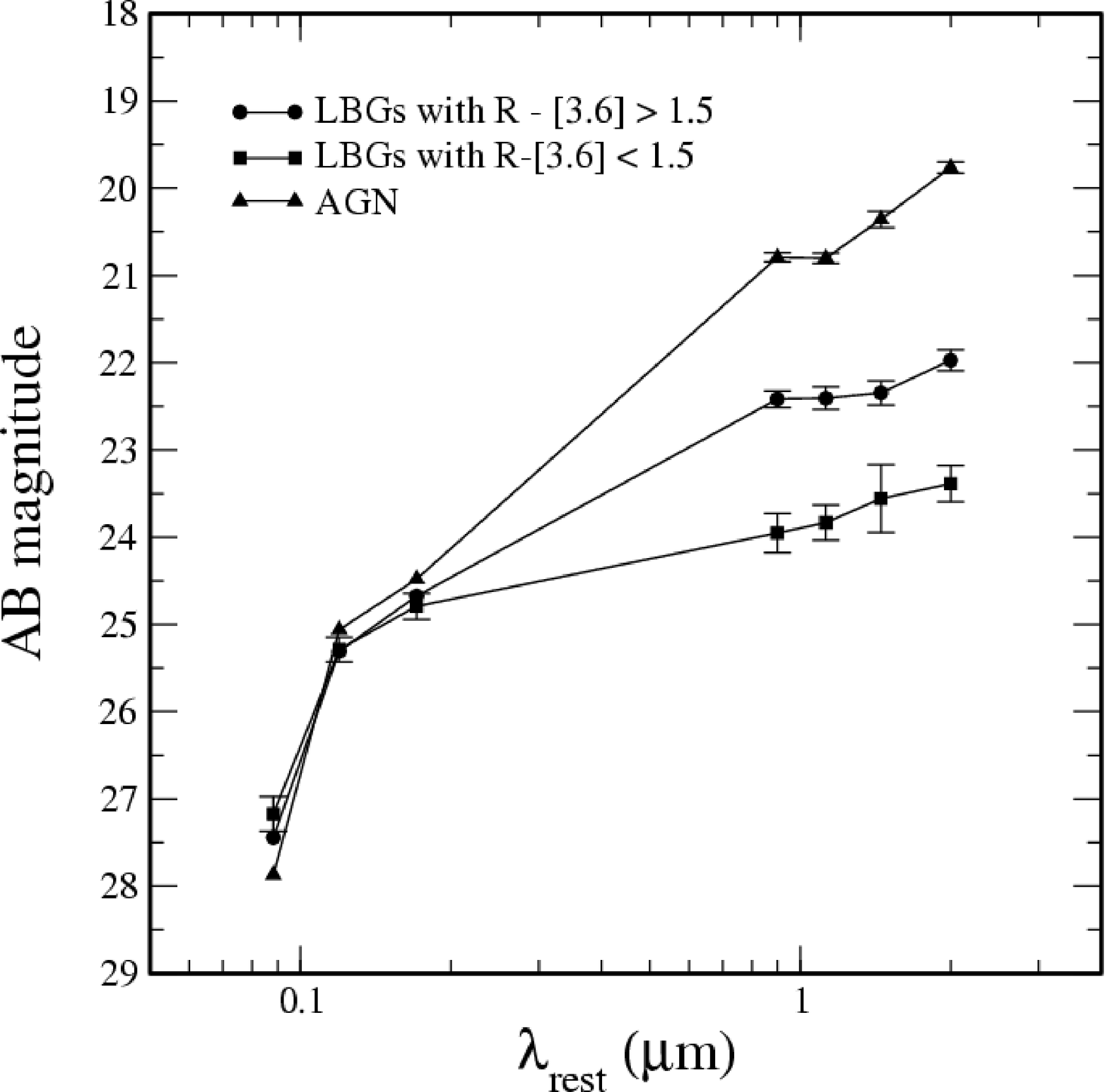}\\
\caption{\small{Average SEDs for AGNs (triangles), LBGs-z (squares) with $R-[3.6]<1.5$ and LBGs-z with $R-[3.6]>1.5$ (circles). We restrict our sample
 to objects of redshift $2.9<z<3.1$.}}
\label{fig:F11}
\end{figure}  
\subsection{The IRAC 8$\mu$m sample}
From the whole LBG sample detected in 3.6$\mu$m and 4.5$\mu$m LBGs, about $\sim$34$\%$ are detected 
in longer wavelengths. This is what we call the 
the sample of the 8$\mu$m LBGs. In total 8$\mu$m counterparts were detected in 152 LBGs and the detection rate is significantly 
lower when compared to that at 3.6$\mu$m and 4.5$\mu$m. The question that should be  answered is how
 the 8$\mu$m LBGs are distributed between the R$-$[3.6] $<$ 1.5 and R$-$[3.6] $>$ 1.5 LBGs and how 
this distribution is affected by the different depths of observations.    
Again, useful conclusion can be derived from the comparison of the two 
sets of fields (HDFN-the rest). The detection rate of 8$\mu$m LBGs in HDFN is $\sim$50$\%$ while for the rest fields falls 
dramatically to $\sim$17$\%$. A simple assumption would be that in the shallower fields
  the 8$\mu$m sample is under-represented by $\sim$33$\%$. As shown in Figure 12, all LBGs in HDFN 
with [3.6]$<$23 are detected at 8.0$\mu$m while for the shallower fields the number of LBGs lacking 8$\mu$m counterpart at 
[3.6]$\sim$23 are comparable with those detected at that band. We can therefore assume that the LBGs in the shallower fields
 having [3.6]$<$23 but not detected at 8$\mu$m do have an 8$\mu$m counterpart but the observation was not 
 deep enough to be detected. This is the minimum estimation of the undetected LBGs at 8$\mu$m,  
as the fraction of detected/undetected at 8$\mu$m remains higher for HDFN for the whole range of [3.6] magnitudes.
\begin{figure}
\centering
\includegraphics[width=8cm,height=8cm,angle=-90]{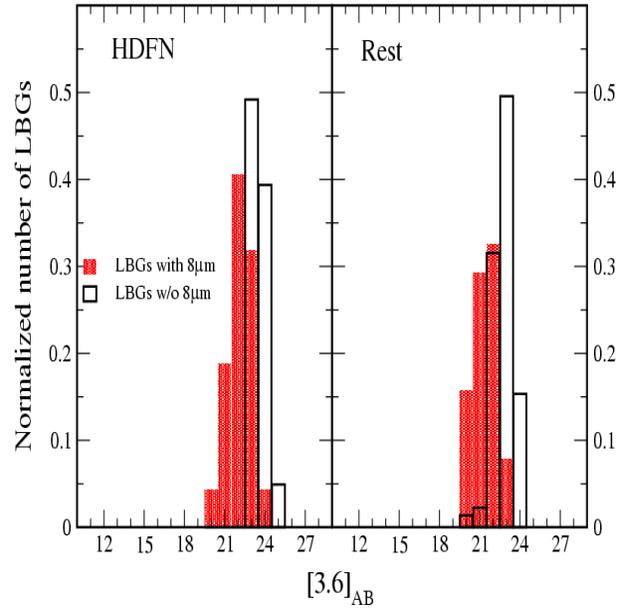}\\
\caption{\small{Normalised [3.6] magnitude  distribution of detected LBGs for LBGs with and without 8$\mu$m detection (red and black bars respectively). The left panel shows the distribution for HDFN, while the left the distribution for the rest of the fields.}}
\label{fig:F12}
\end{figure}
 
This non-detection of 8$\mu$m mainly affects the R$-$[3.6] $>$ 1.5 population of the sample and this is shown clearly in Figure 13.
According to the previous analysis, there are $\sim$40 LBGs in the shallower fields with R-[3.6] $>$ 1.5 that should have been detected
at 8$\mu$m, while for the R$-$[3.6]$<$1.5 the missing LBGs are $\sim$18. If we regard those LBGs as detected at 8$\mu$m, 
we find that that the median 
$<R-[3.6]>$ colour is 1.81$\pm0.164$ with median [3.6] value 22.47$\pm$0.059, while for those 
without 8$\mu$m counterpart is 1.09 $\pm0.26$ with [3.6] median 23.58 $\pm$0.187 (Figure 14).
\begin{figure}
\centering
\includegraphics[width=12cm,height=8cm,angle=-90]{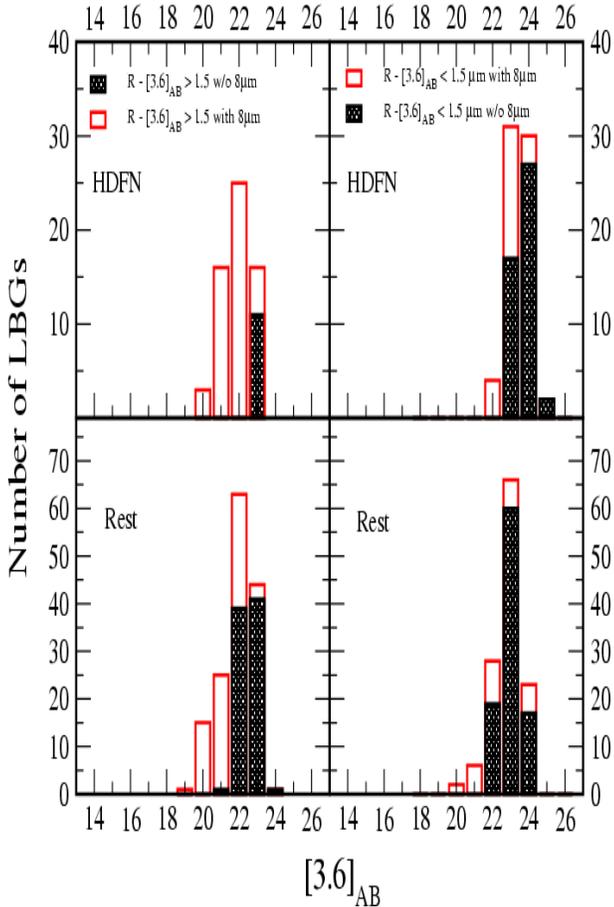}\\
\caption{\small{(Upper left) This figure shows how the $''$red$''$ LBGs, with and without 8$\mu$m detection 
(red and black bars respectively) are distributed in [3.6] magnitude bins. (Bottom left) Same as the upper left panel, but for the rest of the fields. (Upper right) Same as the upper left panel but for the $''$blue$''$ LBGs detected in HDFN. (Bottom right) Same as the upper right panel but for the rest of the fields.}}
\label{fig:F13}
\end{figure}

The significance of the 8$\mu$m sample is that for $z\sim$3, 8$\mu$m correspond to K rest-frame, sensitive to the bulk of the 
stellar emission
of a galaxy
and not only to the young population of a recent star-forming event.
In the first study of \textsl{Spitzer} detection of LBGs-z, Rigopoulou 
et al. 2006 suggests that LBGs-z detected in 8$\mu$m
band are dustier,
more massive and relatively older than those with no detection. But 
her sample was small, covering LBGs-z from the EGS 
(Extended Groth Strip) field. To put the Rigopoulou results on a secure statistical footing,
 in a forthcoming paper we use the current sample to constrain the
 physical properties of the LBGs-z  and investigate the
origin of the 8$\mu$m emission.

\begin{figure}
\includegraphics[width=10cm,height=8cm,angle=-90]{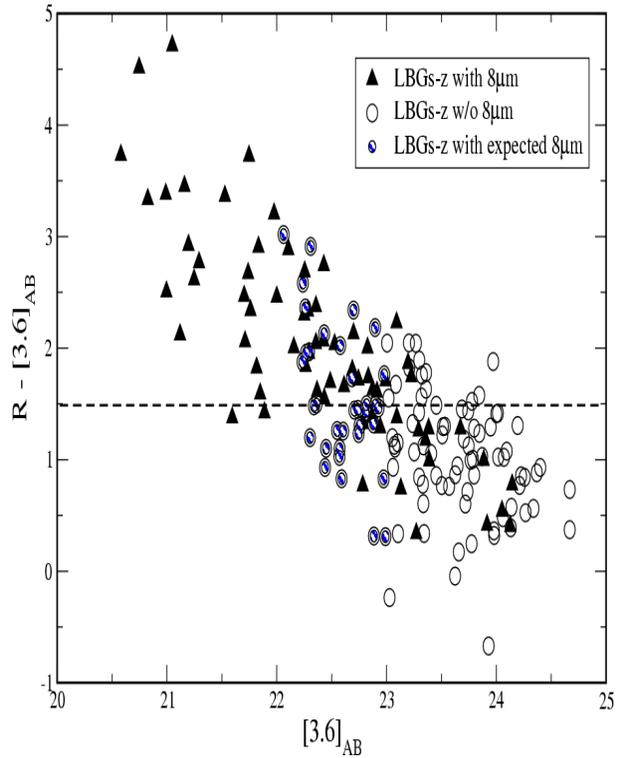}\\
\caption{\small{LBGs-z with 8.0$\mu$m counterparts (triangles) exhibit redder $R-[3.6]$ colours compared 
to those undetected 
in 8.0$\mu$m (empty circles). Shaded circles represent the LBGs of the fields with 
shallow depth that were not detected at 8$\mu$m but we expect that with an exposure time similar 
to that of the HDFN would had been shown up at this band. If those LBGs are considered as detected at 8$\mu$m, 
the mean value of LBGs with 8.0$\mu$m counterpart is 1.81 ($\pm0.164$) while for those not detected 
at 8.0$\mu$m is 1.09 ($\pm0.26$) respectively. The horizontal dashed line corresponds to R$-$[3.6]$=$1.5 }}
\label{fig:F14}
\end{figure}

\subsection{Infrared Colours of LBGs}
As discussed before, LBGs exhibit a much wider range of flux densities in the IRAC bands than in the rest-UV. 
Figure 8 shows that the range 
of 3.6$\mu$m flux densities for the LBGs spans 6 mags, compared to only 3 mag for the range of rest UV-band flux density. To 
investigate the infrared colours of the LBGs, stellar synthesis population models can be employed. 
Figure 15 compares the observed LBG colours with those predicted by two simple stellar 
population synthesis models 
generated with the new Charlot \& Bruzual code that includes an observationally calibrated AGB phase (CB07 private communication).
We considered both single burst and constant star formation models, assuming solar metallicity and Salpeter IMF. 
These two models correspond to two extreme cases of a model having an 
exponentially decaying star formation rate, ${\Psi}(t)=e^{-t/{\tau}}$ where ${\tau}=0.1$  is the single-burst model 
and ${\tau}={\infty}$ is the constant star formation model. The model SEDs were generated for dust free and E(B $-$ V)= 0.3 and for 
a grid of ages, ranging from 1 Myrs to 2 Gyrs. They were then redshifted at z$\sim$3 matching the predicted [3.6]$-$[4.5] and 
R $-$ [3.6] colours of star forming galaxies. Although colour--colour diagrams are not the best way to constrain the properties 
of the stellar population s of a given galaxy class as one would have more free parameters than data points, we note 
that a combination of 
the two models with varying amounts 
of dust attenuation could reproduce the majority of the LBGs (with spectroscopic redshift) colours. This plot also shows 
that most of the LBGs without spectroscopic redshift, 
do have colours that are consistent 
with a galaxy at z$\sim$3, while stars can be screened out from their blue [3.6]$-$[4.5] colours. A more detailed study 
to estimate the stellar masses and the other physical properties of LBGs is currently undertaken based on model SEDs fitting 
(Magdis et al. 2008 in preparation). 
\begin{figure}
\centering
\includegraphics[width=10cm,height=8cm,angle=-90]{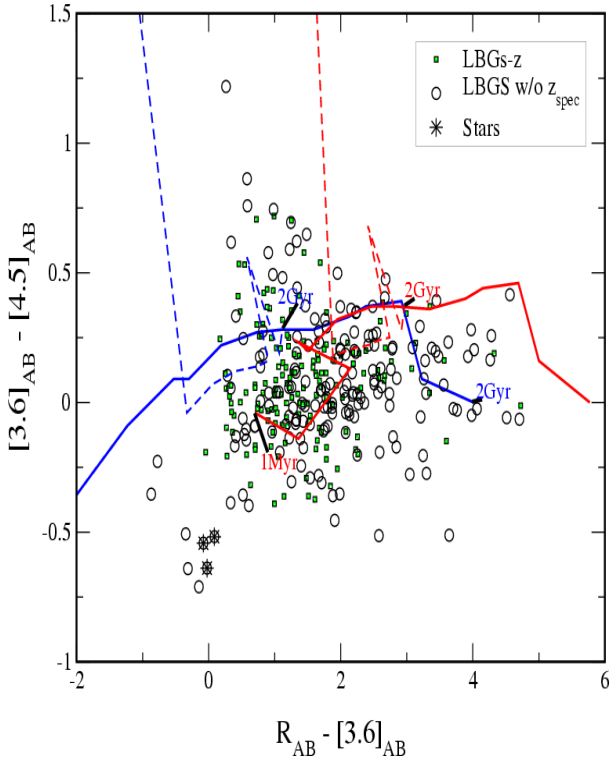}\\
\caption{\small{Observed frame [3.6]$-$[4.5]  vs. R$-$[3.6] colour-colour diagram for all LBGs detected with IRAC. Green squares represent LBGs with spectroscopic redshifts, circles represent LBGs without spectroscopic redshift, while asterisks represent stars.
 The lines show colours from stellar synthesis models based on CB07: solid lines are the single-burst models, and dashed 
lines are the constant star formation models. The blue lines are dust-free models, and the red lines are models with 
E(B $-$ V) = 0.3 and a Calzetti et al. (2000) reddening law. The single-burst models run from 1 Myr at the blue end of the line for 
the reddened model, off the plot for the dust-free model to 2 Gyr as indicated at the red end of the dashed lines. 
The constant star formation models, run from 100Myrs, off the plot for both dust-free and reddened model, to 2Gyr 
as indicated in the red end of the solid lines. Overall the set of models spans nearly the entire colour range of the data. }}
\label{fig:F15}
\end{figure}

\subsection{Energy Source in LBGs}
The IRAC colours can also be used as an indicator, to separate lunimous z$\sim$3 AGN dominated objects 
from normal star-forming galaxies. The fraction of LBGs in our sample with spectroscopically identified AGN is 8/615.
Figure 16
shows that AGN (filled squares) occupy a distinct region in the [4.5]$-$[8.0] over [8.0] 
colour-magnitude plot when compared to LBGs. AGNs are
brighter in 8$\mu$m and exhibit redder [4.5]$-$[8.0] colours. The average [4.5]$-$[8.0] colour 
 for AGNs is 1.22 $\pm0.087$ and for star-forming galaxies is 0.07 $\pm0.22$ while the average
[8.0] is 20.09 $\pm0.04$ and 22.32 $\pm0.11$ respectively. Although most LBGs exhibit similar [4.5]$-$[8.0] colours, 
those with [4.5]$-$[8.0]$<$$-$0.5 must represent a few really young LBGs with blue rest frame J-K colours, while those with 
[4.5]$-$[8.0]$\geq$1.0  must represent the Infared Luminous Lyman Break Galaxies (Huang et al. 2005) with 24$\mu$m detection as they are bright at 8$\mu$m.  
The physical reason 
behind this diagnostic tool, is discussed by several authors (Ward et al. 1987, Elvis et al. 1994, Ivison et
al. 2004. The
SED of an AGN rises with a constant slope at the 0.1--10$\mu$m rest frame interval,
 while for a star-forming galaxy the SED is rather
flat between 
rest 1--3$\mu$m and then rises steeply towards longer wavelengths. 
IRAC bands at z$\sim$3, correspond to rest-near-infrared
(i.e., 0.9--2$\mu$m), so we expect that that AGN should exhibit redder 
[4.5]$-$[8.0] colours than young star-forming galaxies. Also,
an AGN dominated object is expected to have warmer dust and therefore 
having brighter 8$\mu$m counterpart.
Athough a number of works are now showing that the Spitzer selection techniques do miss a large
population of the X-ray selected population of AGN at intermediate X-ray
luminosities (that is, AGN with $L_{x}$$<$$\sim$$10^{44}$erg/s) (e.g. Rigby et al. 2006, Barmby et al. 2006), Reddy et al. 2006 showed that compilation of multiwavelength data for 11 AGNs in HDFN indicates that optical and Spitzer data are able 
to  efficiently (in terms of integration time) select high redshift (z$\sim$3), luminous AGNs. Optical spectra 
and Spitzer and Chandra data are all required to fully account for the census of AGNs at high redshifts.
\begin{figure}
\centering
\includegraphics[width=10cm,height=8cm,angle=-90]{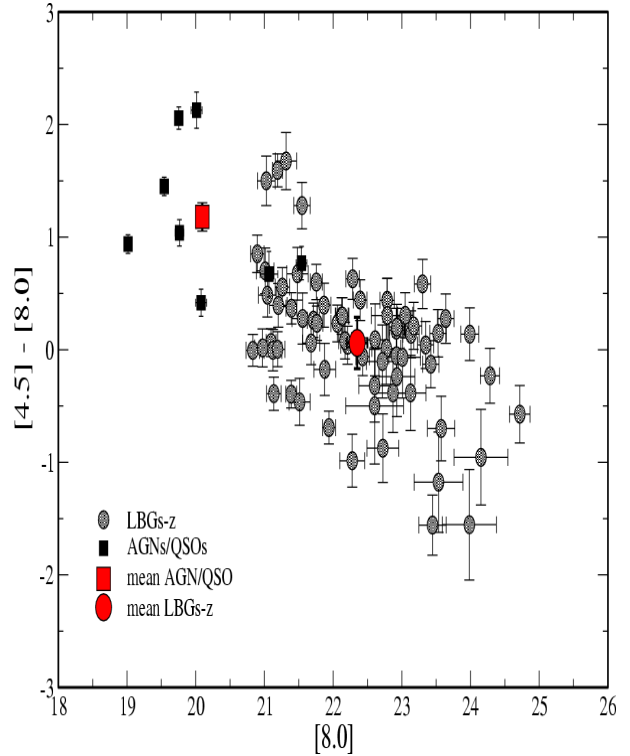}\\
\caption{\small{Colour - Magnitude diagram of [4.5]$-$[8.0] vs. [8.0]. 
A colour-magnitude diagnostic plot for
 weeding out AGNs. AGNs (black squares) occupy a distinct area in that plot having redder [4.5]$-$[8.0] colours and
 being brighter in [8.0] band when compared to the LBG population (shaded circles). 
The red square and circle represent the average colour of AGNs and LBGs--z, respectively 
with AGNs having average [4.5]$-$[8.0] colour of 1.22 ($\pm0.087$) and LBGs--z 0.07 ($\pm0.22$)}}
\label{fig:F16}
\end{figure} 

\section{Summary and Conclusions}
Through the photometric analysis of 6 fields  covered by all four IRAC bands on board Spitzer our conclusions are as follows:

\begin{enumerate}
\item The excellent agreement of the number counts between all the fields in the bright end and between observations of equal exposure time in the faint end, shows 
that our photometric technique and source extraction has treated all fields in the same manner. 

\item Out of $\sim700$ LBGs that were covered by our data, 443, 448, 137 and 152 LBGs were identified at 3.6$\mu$m, 4.5$\mu$m, 5.8$\mu$m, 8.0$\mu$m IRAC bands respectively, creating the largest existing rest-near-infrared sample of high-redshift galaxies. 

\item The SED of the LBGs were expanded to NIR and show that the near-infrared colours of the population spans over 6 magnitudes. 
The addition of IRAC bands reveals for the first time that LBGs display a variety of colours and their rest-near-infrared properties are rather inhomogeneous, ranging from : 
\begin{itemize}
\item Those that are bright in IRAC bands and exhibit $R-[3.6]>1.5$ colours with steeply rising SEDs towards
    longer wavelengths  to
\item Those that are faint or not detected at all in IRAC bands with 
    $R-[3.6]<1.5$ colour whose SEDs are rather flat from the
far-UV to the NIR with marginal IRAC detections.
\end{itemize}

\item Out of the whole sample, $\sim20$$\%$ of the LBGs are detected at 8.0$\mu$m. We refer to them as the 8.0$\mu$m sample of LBGs
(It is equivelent to a rest frame K--selected sample). 
Those LBGs tend to have redder R$-$[3.6] colours when compared to the rest population with median values of 1.81 ($\pm0.16$) 
and 1.09 ($\pm0.26$) respectively.

\item The infrared colours of LBGs are consistent with those of z$\sim3$ galaxies and indicating that their SEDs are can be fitted with various 
stellar synthesis population models. The mid-infrared 
properties of the LBG (i.e., masses, dust, age, link to other z$\sim3$ galaxy populations) will be presented 
in detail in a forthcoming paper as well as the full photometric catalogues. 

\item Based on on results for a few z$\sim$3 optically identified AGN, IRAC 8$\mu$m band can be used as a diagnostic tool to separate luminous, high z, AGN dominated objects from star--forming galaxies with AGNs being 
brighter in [8.0] band when compared to the LBG population.

\end{enumerate} 
This work is based on observations made
with the Spitzer Space Telescope, which is operated by the Jet
Propulsion Laboratory, California Institute of Technology under a
contract with NASA. Support for this work was provided by NASA through
an award issued by JPL/Caltech.

\end{document}